\documentstyle{article}
\begin{document}
\title{X-Ray Edge Spectra From Sea-Bosons-I}
\author{Girish S. Setlur}
\maketitle

\begin{abstract}
The well-studied phenomenon of X-ray edge singularities is
revisited using the sea-boson approach that has recently been
placed on a rigorous footing. We are able to reproduce the
well-known result namely, Mahan's power law divergences. Unlike
the work of Schotte and Schotte, no linearization of the bare
fermion dispersion is needed, which, by their own admission, is a
source of some difficulty. Our approach also brings out some
differences between the different dimensions which is not present
in their work. Finally, our work also allows for easy
generalization to potentials more realistic than the simple
delta-function used commonly in the literature.
\end{abstract}

\section{Introduction}

The X-ray edge problem is by now extremely well understood. The
work of Anderson\cite{And},Mahan\cite{Mahan} and Nozi\'{e}res et.
al. \cite{Noz}, are the early pioneering works. Schotte and
Schotte\cite{Sch} were the first to apply the technique of
bosonization to this problem. More recently, E\ss ler  and Frahm
\cite{Ess} have used the exact Bethe ansatz solution to the
Hubbard model in 1D, in their study of the X-ray edge spectra. The
striking feature of this phenomenon is the power law divergence of
the absorption close to and above a certain threshold. This is
known to come about as a result of the filled Fermi sea being
strongly perturbed by the emergence of the core hole, thereby a
significant departure from the case when the hole is absent may be
expected. The precise nature of this departure requires detailed
analysis using the techniques of Many-Body theory to which we now
turn. The sea-boson technique is so versatile that it allows for
the study of this problem with varying degrees of sophistication
and rigor. The simplest version is the one found most commonly
discussed in the literature. It involves assuming that a core hole
appears instantaneously when x-ray photons are absorbed, with a
constant (time independent) interaction between the electrons in
the conduction band and the core hole. Clearly, this is an over
simplification. A more desirable model involves a time-dependent
build up of the interaction between the electron and the hole that
reaches a final screened equilibrium form determined dynamically.
This sort of build up of screening has been discussed by us in the
past in a different context\cite{Setlur2}. In this article, (with
Roman numeral I) we consider the simplest version as outlined in
the work by Schotte and Schotte\cite{Sch}. In a future
publication, we intend treating the more realistic version.

\section{ The Hamiltonian }

The hamiltonian in the Fermi language may be written down as
follows. We try to follow the notation of Schotte and
Schotte\cite{Sch}. In what follows, $ c_{ {\bf{k}} } $ annihilates
a conduction
 electron with momentum $ {\bf{k}} $ and $ b $ annihilates a core electron
 (not a hole).
\begin{equation}
H = \sum_{ {\bf{k}} }\epsilon_{ {\bf{k}} }c^{\dagger}_{ {\bf{k}}
}c_{ {\bf{k}} } - \frac{1}{N} \sum_{ {\bf{k}} \neq {\bf{k}}^{'} }
v({\bf{k}}-{\bf{k}}^{'})  c^{\dagger}_{ {\bf{k}} }c_{ {\bf{k}}^{'}
}b\mbox{         }b^{\dagger}
 + E_{0}b^{\dagger}b
 \label{HAMIL}
\end{equation}
 Being fermions, we expect the operators to obey,
$ \{c_{ {\bf{k}} },c^{\dagger}_{ {\bf{k}} }\} = 1 $ and $
\{b,b^{\dagger}\} = 1 $
 and all other anticommutators involving any two of these
 operators are zero.
 Here $ \epsilon_{ {\bf{k}} } = k^{2}/(2m) $ is the kinetic energy of
 the conduction electrons and $ m > 0 $ is the band mass of the
 electron. We use units such that $ \hbar = 1 $.
 Also $ E_{0} < 0 $ is the single nondegenerate core energy level, which is occupied
 by the core electron, unless excited.
 We assume that the intra-band interactions may be ignored,
 since the important physics is contained in the inter-band interaction.
We assume, just as all others do, that the interaction between the
core electron and the conduction electron is of the delta-function
type which makes the analysis simpler. However, one of the main
selling points of our technique is the ease with which one may
study more realistic potentials in three dimensions as well as one
and two dimensions.
 In the simplest case we set $ v_{ {\bf{q}} } = v_{0} $.
  The way Eq.(~\ref{HAMIL}) is
written, it is clear that $ v_{0} > 0 $. We now proceed to solve
for the absorption spectra using Fermi's Golden rule.

\section{Absorption Spectra from Fermi's Golden Rule}

We borrow from the work of Schotte and Schotte\cite{Sch} who have
computed the absorption from Fermi's Golden rule.
\begin{equation}
W(\omega) = 2\pi\mbox{    }w^{2}\sum_{f}\left| \langle f_{n+1} |
\frac{1}{\sqrt{N}}\sum_{ {\bf{k}} }c^{\dagger}_{ {\bf{k}} } |
i_{n} \rangle \right|^{2} \mbox{         }
\delta(E_{i}-E_{f}+\omega)
\end{equation}
The initial state is $ |i_{n}\rangle $ consists of a
noninteracting filled Fermi sea with $ n $ particles. It is the
ground state of the hamiltonian below. We have set
 $ b^{\dagger}b = 1 $ and $ bb^{\dagger} = 0 $.
\begin{equation}
H_{i} = \sum_{ {\bf{k}} }\epsilon_{ {\bf{k}} }c^{\dagger}_{
{\bf{k}} } c_{ {\bf{k}} } + E_{0}\label{HAMILI}
\end{equation}
The final states are the eigenstate of the hamiltonian that is
shifted corresponding to the emergence of a hole in the core state
and the core electron being promoted to the conduction band. The
final state is an eigenstate of a set of $ n + 1 $ electrons
interacting with a core hole. Here we have to set $ b^{\dagger}b =
0 $ and $ bb^{\dagger} = 1 $.
\begin{equation}
H_{f} = \sum_{ {\bf{k}} }\epsilon_{ {\bf{k}} }c^{\dagger}_{
{\bf{k}} } c_{ {\bf{k}} } - \sum_{ {\bf{q}} \neq 0 }\frac{v_{
{\bf{q}} }}{N}\rho_{ {\bf{q}} }
 \label{HAMILF}
\end{equation}
Define,
\begin{equation}
a = \frac{1}{\sqrt{N}}\sum_{ {\bf{k}} }c_{ {\bf{k}} }
\end{equation}
and,
\begin{equation}
{\mathcal{F}}(t) = \langle i | e^{i\mbox{    }t\mbox{    }H_{i}}
\mbox{     }a\mbox{      } e^{-i\mbox{    }t\mbox{    }H_{f}}
\mbox{          }a^{\dagger}\mbox{           }|i\rangle
\end{equation}
It can be shown quite easily that,
\begin{equation}
W(\omega) = 2\pi\mbox{   }w^{2}\mbox{         } Im \frac{i}{\pi}
\int^{\infty}_{0} e^{i\omega t}\mbox{   } {\mathcal{F}}(t)\mbox{
}dt
\end{equation}
 The problem is now well-posed
and we proceed to solve it using the sea-boson method. In our
earlier work\cite{Setlur1}, we have introduced the sea-boson
method and made it rigorous in another recent
preprint\cite{Setlur1}. The sea-displacement operators
 $ A_{ {\bf{k}} }({\bf{q}}) $ are introduced and postulated to be
 exact bosons in these sense below.
\begin{equation}
[A_{ {\bf{k}} }({\bf{q}}),A_{ {\bf{k}}^{'} }({\bf{q}}^{'})] =
0\mbox{         };\mbox{            } [A_{ {\bf{k}}
}({\bf{q}}),A^{\dagger}_{ {\bf{k}}^{'} }({\bf{q}}^{'})] = \delta_{
{\bf{k}}, {\bf{k}}^{'} }\delta_{ {\bf{q}}, {\bf{q}}^{'} }
n_{F}({\bf{k}}-{\bf{q}}/2)(1-n_{F}({\bf{k}}+{\bf{q}}/2))
\end{equation}
The Fermi bilinears are related to these objects as follows( $
{\bf{q}} \neq 0 $ ).
\begin{equation}
c^{\dagger}_{ {\bf{k}} }c_{ {\bf{k}} } = n_{F}({\bf{k}})
 + \sum_{ {\bf{q}} }A^{\dagger}_{ {\bf{k}}-{\bf{q}}/2 }({\bf{q}})
A_{ {\bf{k}}-{\bf{q}}/2 }({\bf{q}}) -
 \sum_{ {\bf{q}} }A^{\dagger}_{ {\bf{k}}+{\bf{q}}/2 }({\bf{q}})
A_{ {\bf{k}}+{\bf{q}}/2 }({\bf{q}})
\end{equation}
\begin{equation}
c^{\dagger}_{ {\bf{k}}+{\bf{q}}/2 }c_{ {\bf{k}}-{\bf{q}}/2 } = A_{
{\bf{k}} }(-{\bf{q}}) + A^{\dagger}_{ {\bf{k}} }({\bf{q}})
\end{equation}
In the sea-boson language, we have,
\begin{equation}
H_{i} = \sum_{ {\bf{k}},{\bf{q}} }\omega_{ {\bf{k}}
}({\bf{q}})A^{\dagger}_{ {\bf{k}} }({\bf{q}})A_{ {\bf{k}}
}({\bf{q}}) + E_{0}
\end{equation}
\begin{equation}
H_{f} = \sum_{ {\bf{k}},{\bf{q}} }\omega_{ {\bf{k}}
}({\bf{q}})A^{\dagger}_{ {\bf{k}} }({\bf{q}})A_{ {\bf{k}}
}({\bf{q}}) -\sum_{ {\bf{k}},{\bf{q}} \neq 0 } \frac{v_{ {\bf{q}}
}}{N}\left[ A_{ {\bf{k}} }(-{\bf{q}}) + A^{\dagger}_{ {\bf{k}}
}({\bf{q}}) \right]
\end{equation}
Here, $ \omega_{ {\bf{k}} }({\bf{q}}) = {\bf{k.q}}/m $. Following
Schotte and Schotte we introduce a unitary transformation that
transforms $ H_{i} $ into $ H_{f} $
\begin{equation}
H_{f} = S^{\dagger}\mbox{  }H_{i}\mbox{  }S - E^{'}_{0}
\end{equation}
\begin{equation}
S = exp \left[ \sum_{ {\bf{k}}{\bf{q}} }\frac{v_{ {\bf{q}} }}{N}
\frac{A_{ {\bf{k}} }({\bf{q}}) - A^{\dagger}_{ {\bf{k}}
}({\bf{q}})}{\omega_{ {\bf{k}} }({\bf{q}})} \right]
\end{equation}
where,
\begin{equation}
-E^{'}_{0} = -E_{0} -\sum_{ {\bf{k}}{\bf{q}} } \frac{v^{2}_{
{\bf{q}} }}{N^{2}}\frac{ \Lambda_{ {\bf{k}} }(-{\bf{q}}) }{
\omega_{ {\bf{k}} }({\bf{q}}) }
\end{equation}
and $ \Lambda_{ {\bf{k}} }(-{\bf{q}}) =
n_{F}({\bf{k}}-{\bf{q}}/2)(1-n_{F}({\bf{k}}+{\bf{q}}/2)) $. The
field operator may also be expressed in terms of the sea-bosons as
we have showed earlier\cite{Setlur1}.
\begin{equation}
\psi({\bf{x}}) = e^{-i\mbox{   }\Pi({\bf{x}})}e^{i\mbox{
}\Phi([\rho];{\bf{x}}) }\sqrt{ \rho_{0} }
\end{equation}
\begin{equation}
\Pi({\bf{x}}) = \sum_{ {\bf{q}}
}e^{i{\bf{q.x}}}\left(\frac{1}{-2\mbox{    }i\mbox{     }
N\epsilon_{ {\bf{q}} }}\right) \sum_{ {\bf{k}} } \omega_{ {\bf{k}}
}({\bf{q}})\left[ A_{ {\bf{k}} }({\bf{q}}) + A^{\dagger}_{
{\bf{k}} }(-{\bf{q}}) \right]
\end{equation}
\begin{equation}
\Phi([\rho];{\bf{x}}) = \sum_{ {\bf{q}} }e^{-i{\bf{q.x}} }\mbox{ }
U_{ {\bf{q}} }\mbox{  }\sum_{ {\bf{k}} }\left[ A_{ {\bf{k}}
}(-{\bf{q}}) + A^{\dagger}_{ {\bf{k}} }({\bf{q}}) \right]
\end{equation}
Here $ U_{ {\bf{q}} } $ is a real number whose exact form is not
important.
 It is clear from the definition that $ a =
\psi({\bf{0}})/\sqrt{\rho_{0}} $, where $ \rho_{0} = N/V $ is the
density of conduction electrons. This means,
\begin{equation}
W(\omega) = 2\pi\mbox{    }w^{2}\mbox{   }Im\frac{i}{\pi}
\int^{\infty}_{0}dt\mbox{          }e^{i(\omega+E^{'}_{0})t}
\mbox{      }\langle i | B^{\dagger}(t)\mbox{  }B(0)| i \rangle
\end{equation}
Where,
\begin{equation}
B^{\dagger}(t) = e^{i\mbox{    }t\mbox{    }H_{i}}\mbox{ }a\mbox{
}S^{\dagger}\mbox{   }e^{-i\mbox{    }t\mbox{    }H_{i}}
\end{equation}
It can be shown after some algebra,
\begin{equation}
\langle i | B^{\dagger}(t)\mbox{  }B(0)| i \rangle = exp\left[
\sum_{ {\bf{k}}{\bf{q}} } |T_{ {\bf{k}} }({\bf{q}})|^{2} \mbox{ }
\Lambda_{ {\bf{k}} }(-{\bf{q}}) \mbox{ } \left( e^{-i\mbox{ }
\omega_{ {\bf{k}} }({\bf{q}})\mbox{   }t} - 1 \right) \right]
\label{BDAGB}
\end{equation}
\begin{equation}
T_{ {\bf{k}} }({\bf{q}}) = \frac{ \omega_{ {\bf{k}} }({\bf{q}}) }{
2\mbox{   }N\mbox{   }\epsilon_{ {\bf{q}} } } -\frac{v_{ {\bf{q}}
}}{N\mbox{ }\omega_{ {\bf{k}} }({\bf{q}})} + i \mbox{      }U_{
{\bf{q}} }
\end{equation}
The quantity $ U_{ {\bf{q}} } $ may be eliminated by making
contact with free theory. Using the approach outlined in our
earlier work\cite{Setlur1}, we multiply and divide
Eq.(~\ref{BDAGB}) by the noninteracting version of the correlation
function, in the numerator, we use the expression deduced using
Fermi algebra and in the denominator, use the form in
Eq.(~\ref{BDAGB}) but with $ v_{0} = 0 $. Thus we have,
\begin{equation}
\langle i | B^{\dagger}(t)\mbox{  }B(0)| i \rangle = exp\left[
\sum_{ {\bf{k}}{\bf{q}} } |t_{ {\bf{k}} }({\bf{q}})|^{2} \mbox{ }
\Lambda_{ {\bf{k}} }(-{\bf{q}}) \mbox{ } \left( e^{-i\mbox{ }
\omega_{ {\bf{k}} }({\bf{q}})\mbox{   }t} - 1 \right) \right]
\langle i | B^{\dagger}_{0}(t)\mbox{  }B_{0}(0)| i \rangle
\end{equation}
here,
\begin{equation}
|t_{ {\bf{k}} }({\bf{q}})|^{2} = \left( \frac{ \omega_{ {\bf{k}}
}({\bf{q}}) }{ 2\mbox{   }N\mbox{   }\epsilon_{ {\bf{q}} } }
-\frac{v_{ {\bf{q}} }}{N\mbox{    }\omega_{ {\bf{k}} }({\bf{q}})}
\right)^{2} + \mbox{ }U^{2}_{ {\bf{q}} } -\left( \frac{ \omega_{
{\bf{k}} }({\bf{q}}) }{ 2\mbox{   }N\mbox{   }\epsilon_{ {\bf{q}}
} } \right)^{2} - \mbox{ }U^{2}_{ {\bf{q}} }
\end{equation}
Miraculously, the dependence on $ U_{ {\bf{q}} } $ cancels. Also,
the free case is evaluated quite simply as,
\begin{equation}
\langle i | B^{\dagger}_{0}(t)\mbox{  }B_{0}(0)| i \rangle =
\frac{1}{\rho_{0}}\langle
\psi({\bf{0}},t)\psi^{\dagger}({\bf{0}},0)\rangle
 = \frac{1}{N}\sum_{ {\bf{k}} }(1-n_{F}({\bf{k}}))
e^{-i\epsilon_{ {\bf{k}} }t}
\end{equation}
The best way to evaluate these expressions is to work with the
Fourier transform of the exponent with the double summation.
Define,
\begin{equation}
I(\omega) = \sum_{ {\bf{k}}{\bf{q}} }|t_{ {\bf{k}}
}({\bf{q}})|^{2} \Lambda_{ {\bf{k}} }(-{\bf{q}})\delta(\omega -
\frac{ {\bf{k.q}} }{m} )
\end{equation}
This may be rewritten as,
\begin{equation}
I(\omega) = \frac{1}{N}\sum_{ {\bf{q}} }\left( \frac{v^{2}_{
{\bf{q}} }}{\omega^{2}} - \frac{v_{ {\bf{q}} }}{\epsilon_{
{\bf{q}} }} \right)f({\bf{q}},\omega)
\end{equation}
\begin{equation}
f({\bf{q}},\omega) =  \frac{1}{N}\sum_{ {\bf{k}} }\Lambda_{
{\bf{k}} }(-{\bf{q}})\delta(\omega - \frac{ {\bf{k.q}} }{m} )
\end{equation}
It is clear at the
 outset that since $ \Lambda_{ {\bf{k}}}(-{\bf{q}}) = 0 $
 if $ {\bf{k.q}} < 0 $,
 we must have $ f({\bf{q}},\omega) = 0 $ if $ \omega < 0 $.
 Since we are interested in the $ t \rightarrow \infty $
 limit, we have to investigate the $ \omega \rightarrow 0^{+} $ limit
 of these expressions.
%\newpage
\[
f({\bf{q}},\omega) =  \frac{1}{N}\sum_{ {\bf{k}} } \theta(k_{F} -
(k^{2} + q^{2}/4 -m\mbox{      } \omega)^{\frac{1}{2}})
\theta((k^{2} + q^{2}/4 + m\mbox{      }
\omega)^{\frac{1}{2}}-k_{F}     ) \delta(\omega - \frac{
{\bf{k.q}} }{m} )
\]
\[
 =  \frac{(2\pi)^{-d}}{\rho_{0}}\int_{0}^{\infty} \mbox{     }dk\mbox{    } k^{d-1}
  \theta(k_{F} -
(k^{2} + q^{2}/4 -m\mbox{      } \omega)^{\frac{1}{2}})
\theta((k^{2} + q^{2}/4 + m\mbox{      }
\omega)^{\frac{1}{2}}-k_{F})
\]
\[
\times \int d\Omega_{d}\mbox{        } \delta(\omega - \frac{
{\bf{k.q}} }{m} )
\]

\subsection{ Computations in 1D }

Here we choose $ v_{ {\bf{q}} } = v_{0} $. This enables a
comparison with the work of Schotte and Schotte. We may evaluate
in 1D( Since $ k
> 0 $ ),
\[
\int d\Omega_{d=1}\mbox{        } \delta(\omega - \frac{
{\bf{k.q}} }{m} )  =  \mbox{        } \delta(\omega - \frac{
|k||q| }{m} )
\]
\[
f(q,\omega) = \theta(\omega)\mbox{       }\left(
\frac{m}{2k_{F}|q|} \right)\theta(k_{F} - (\frac{ m^{2}\omega^{2}
}{q^{2}}+\frac{ q^{2} }{ 4}-m\mbox{ }\omega)^{\frac{1}{2}})\mbox{
} \theta((\frac{ m^{2}\omega^{2}}{q^{2}}+\frac{ q^{2} }{4}+m\mbox{
}\omega)^{\frac{1}{2}}-k_{F})
\]
Unfortunately, $ f(q,\omega) $ is not an analytic function of $
\omega $ and it is not possible to expand in powers of $ \omega $.
We will have to evaluate the sums $ \sum_{q} f(q,\omega) $ and
 $ \sum_{q} f(q,\omega) /\epsilon_{q} $. For this we have to
 identify the regions in $ q \in [0,\infty) $ for which the
 product of the theta functions is equal to unity. This is done by
 equating the arguments of the theta functions to zero and
 computing the roots. There are four roots, each pair defines an
 interval of integration. The intervals in question are
  $ q \in [ \frac{m\omega}{k_{F}}(1-\frac{m\omega}{k^{2}_{F}}),
  \frac{m\omega}{k_{F}}(1+\frac{m\omega}{k^{2}_{F}})] $ and
 $ q \in [ 2k_{F} - \frac{m\omega}{k_{F}},2k_{F} + \frac{m\omega}{k_{F}}]
 $.
 Since in order for our theory to be consistent\cite{Setlur1}, we have to assume
 $ |q| << k_{F} $, we shall ignore the second interval.
\begin{equation}
\frac{1}{N}\sum_{q}\mbox{  }f(q,\omega) = \omega\mbox{ }
\theta(\omega)\mbox{ }\frac{m^{2}}{k_{F}^{4}}
\end{equation}
\begin{equation}
\frac{1}{N}\sum_{q}\mbox{  }f(q,\omega)/\epsilon_{q} = \mbox{ }
\theta(\omega)\mbox{ }\frac{2m}{k^{2}_{F}}\frac{1}{\omega}
\end{equation}
Therefore,
\begin{equation}
I(\omega) = \left[ \frac{\delta^{2}}{\pi^{2}} - 2
\frac{\delta}{\pi} \right] \frac{ \theta(\omega) }{\omega}
\end{equation}
where,
\begin{equation}
\frac{\delta}{\pi} = \frac{ m \mbox{   }v_{0} }{k^{2}_{F}}
\end{equation}
A similar approach is called for when evaluating the corresponding
coefficient.
\begin{equation}
{\mathcal{G}}_{0}(\omega) =
\frac{1}{k_{F}}\int_{k_{F}}^{\infty}dk\mbox{ } \delta(\omega -
\frac{k^{2}}{2m})
 =  \frac{1}{k_{F}}\frac{\sqrt{2m\mbox{   }\omega}}{m}
 \theta(\omega-\epsilon_{F})
\end{equation}
The Fourier transform may be written as,
\begin{equation}
{\mathcal{G}}_{0}(t) \approx \frac{1}{k_{F}}\frac{\sqrt{2m\mbox{
}\epsilon_{F}}}{m}\frac{ e^{-i\epsilon_{F} t} }{it}
\end{equation}
\begin{equation}
{\mathcal{G}}(t) \propto e^{I(t)} \mbox{     }{\mathcal{G}}_{0}(t)
\end{equation}
\begin{equation}
I(t) = \left[ \frac{\delta^{2}}{\pi^{2}} - 2 \frac{\delta}{\pi}
\right]Log \left( \frac{1}{t} \right)
\end{equation}
In other words,
\begin{equation}
W(\omega) \propto \int^{\infty}_{0} dt\mbox{    } \frac{ cos\left[
(\omega + E^{'}_{0} - \epsilon_{F})t \right] }{ t^{
(1-\delta/\pi)^{2} } } \propto ( \omega + E^{'}_{0} - \epsilon_{F}
)^{ -[2\delta/\pi-(\delta/\pi)^{2}] }
\end{equation}
This is in exact agreement with the result of Schotte and
Schotte\cite{Sch}.

\subsection{Computations in 3D}

Let us now study the screened long-range interaction. This
includes implicitly both the genuinely long-ranged and the
genuinely short-range cases as limiting cases. We shall soon see
that the unscreened interaction leads to a divergent absorption
for all frequencies, hence it is important to screen the
long-range interaction. We choose the simple Thomas-Fermi
screening. In 3D, we have,
\begin{equation}
 f({\bf{q}},\omega)=  \frac{m}{4\pi^{2}\mbox{   }\rho_{0}}
 \left( \frac{ m\mbox{  }\omega }{|q|} \right)
 \theta( \frac{ k_{F} |q| }{m} - \omega )
 \theta(q_{max} - |q|)
\end{equation}
Here we have assumed tacitly, $ \omega << \epsilon_{F} $ and $ |q|
<< k_{F} $. It is clear from the general definition that $
f({\bf{q}},\omega) = 0 $ for $ |q| > 2k_{F} $. However, since our
approach is valid for small $ |q| << k_{F} $,
 we set $ q_{max} = k_{F}/2 $, following the suggestion of Schotte
 and Schotte\cite{Sch}.
 Here we have,
\begin{equation}
I(\omega) = \frac{1}{N}\sum_{ {\bf{q}} }\left( \frac{v^{2}_{
{\bf{q}}  }}{\omega^{2}} - \frac{v_{ {\bf{q}} }}{\epsilon_{
{\bf{q}} }} \right)f({\bf{q}},\omega)
\end{equation}
If we choose $ v_{ {\bf{q}} } = (4 \pi \rho_{0}
e^{2}/\epsilon_{\infty})(q^{2}+q^{2}_{TF})^{-1} $, it is clear
that for $ \omega \rightarrow 0 $ only the positive part in the
above sum contributes just as earlier, and we expect an approach
to zero at the threshold rather than a divergence.
\begin{equation}
I(\omega) =  \frac{ \delta^{2}_{0} }{\omega}
\end{equation}
where $ \delta^{2}_{0} = (\pi a_{B} q_{TF})^{-2} $ and
 $ a^{-1}_{B} = me^{2}/\epsilon_{\infty} $.
 This means,
\begin{equation}
W(\omega) \propto \int^{\infty}_{0}dt \frac{ cos\left[ (\omega +
E^{'}_{0} - \epsilon_{F})t \right] }{t^{(1+\delta_{0}^{2})}}
\propto (\omega + E^{'}_{0} - \epsilon_{F})^{\delta^{2}_{0}}
\end{equation}

\section{Conclusions}

We conclude by examining these results and comparing them with the
results of Schotte and Schotte\cite{Sch}. In 1D, we have found
nearly exact agreement with their results. The only difference is
that the exponent seems to be different somewhat. In our case for
a fixed $ v_{0} $ and $ m $ the exponent scales as $ k^{-2}_{F} $,
whereas in their case it seems to scale as $ k^{-1}_{F} $. This is
not surprising since in the Tomonaga model we have to choose
 $ k_{F} \propto m $ since we take $ k_{F}, m \rightarrow \infty $
 such that $ k_{F}/m = v_{F} < \infty $ in order to make the
 dispersion linear. In Fig. 1 we see the various cases in 3D.
 For strong screening, we have infinite slope and for weak
 screening we have a zero slope and for optimum screening we have
 slope equal to unity.

 This work was supported by the JNCASR under the postdoctoral
 fellowship program. Useful conversations with Prof. H.R.
 Krishnamurthy, Prof. D. Sen, Prof. T.V. Ramakrishnan and
  Prof. B.S. Shastry are gratefully acknowledged.

\begin{figure}
\begin{picture}(60,150)(-15,-25)

\put(1,1){\line(1,0){299}} \put(1,201){\line(1,0){299}}
\put(1,1){\line(0,1){200}} \put(300,1){\line(0,1){200}}

\put(60,165){ $ \delta^{2}_{0} < 1 $ } \put(100,90){ $
\delta^{2}_{0} = 1 $ } \put(200,60){ $ \delta^{2}_{0} > 1 $ }

 \qbezier[300](1,1)(0,200)(300,200) \put(1,1){\line(3,2){299}}
\qbezier[100](1,1)(200,0)(300,200)

\put(150,-15){ $ \omega - \omega_{T} $ } \put(-10,-10){ $ {\bf{0}}
$ }

\put(-40,100){ $ W(\omega)$ }

\end{picture}
\caption{Threshold X-ray Spectra}
\end{figure}
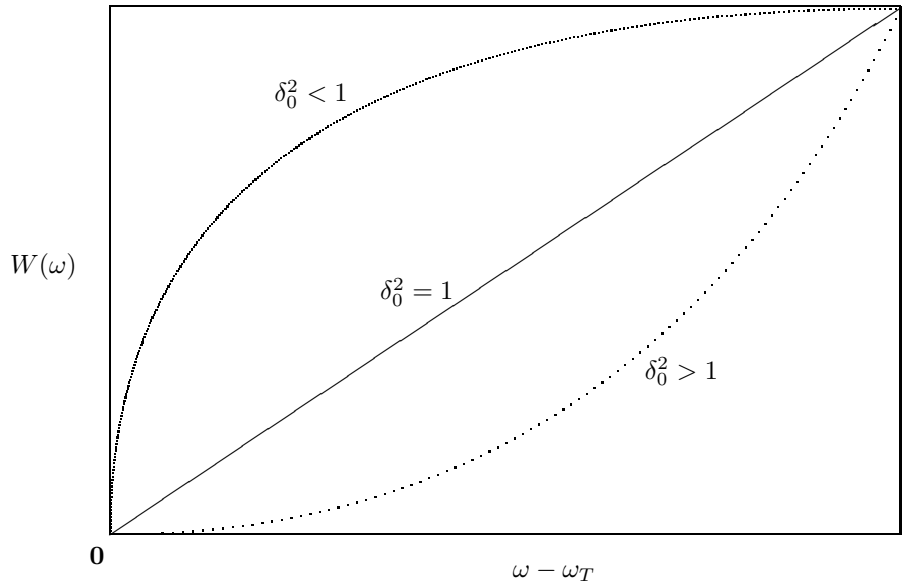

%\begin{figure}
%\include{plot.ps}
%\label{Fig. 1}
%\end{figure}

\end{document}